\documentclass[%
superscriptaddress,
reprint,
 amsmath,amssymb,
prb
]{revtex4-1}
\usepackage{color}
\usepackage{graphicx}
\usepackage{dcolumn}
\usepackage{bm}

\begin{document}

\title{Direct evidence for the existence of heavy quasiparticles in the magnetically ordered phase of CeRhIn$_5$}

\author{Masahiro Haze}
\email{haze.masahiro.2x@kyoto-u.ac.jp}
 \affiliation{Department of Physics, Kyoto University, Sakyo-ku, Kyoto 606-8502, Japan}
 \author{Robert Peters}
 \affiliation{Department of Physics, Kyoto University, Sakyo-ku, Kyoto 606-8502, Japan}
\author{Yohei Torii}
 \affiliation{Department of Physics, Kyoto University, Sakyo-ku, Kyoto 606-8502, Japan}
\author{Tomoka Suematsu}
 \affiliation{Department of Physics, Kyoto University, Sakyo-ku, Kyoto 606-8502, Japan}
\author{Daiki Sano}
 \affiliation{Department of Physics, Kyoto University, Sakyo-ku, Kyoto 606-8502, Japan}
\author{Masahiro Naritsuka}
 \affiliation{Department of Physics, Kyoto University, Sakyo-ku, Kyoto 606-8502, Japan}
\author{Yuichi Kasahara}
 \affiliation{Department of Physics, Kyoto University, Sakyo-ku, Kyoto 606-8502, Japan}
\author{Takasada Shibauchi}
 \affiliation{Department of Advanced Materials Science, The University of Tokyo, Kashiwa, Chiba 277-8561, Japan}
\author{Takahito Terashima}
 \affiliation{Department of Physics, Kyoto University, Sakyo-ku, Kyoto 606-8502, Japan}
\author{Yuji Matsuda}
 \affiliation{Department of Physics, Kyoto University, Sakyo-ku, Kyoto 606-8502, Japan}

\newcommand{\rp}[1]{{\color{red} #1}}
\newcommand{\mh}[1]{{\color{red} #1}}
\newcommand{\cm}[1]{}

\date{\today}

\begin{abstract}
It is a long-standing important issue in heavy fermion physics whether $f$-electrons are itinerant or localized when the magnetic order occurs.  Here we report the {\it in situ} scanning tunneling microscopy observation of the electronic structure in epitaxial thin films of CeRhIn$_5$, a prototypical heavy fermion compound with antiferromagnetic ground state.   The conductance spectra above the N\'eel  temperature $T_N$ clearly resolve the energy gap due to the hybridization between local 4$f$ electrons and conduction bands as well as  the crystal electric field excitations. These structures persist even below $T_N$.  Moreover,  an additional  dip in the conductance spectra develops due to the antiferromagnetic order.   These results provide direct evidence for the presence of itinerant heavy $f$-electrons participating in the Fermi surface even in the magnetically ordered state of CeRhIn$_5$.

\end{abstract}

\pacs{Valid PACS appear here}

\maketitle

\makeatletter
\noindent

\section{introduction}

In heavy fermion compounds, containing 4$f$ (lanthanoid) and 5$f$(actinoid) electrons, $f$-electrons are essentially localized with well defined magnetic moments at high temperature.   As the temperature is lowered, the $f$ electrons begin to delocalize due to the hybridization with the conduction electron band ($c$-$f$ hybridization) and the Kondo  effect.   At yet lower temperatures the $f$-electrons become itinerant, forming a narrow conduction band with heavy effective mass. While the Kondo effect is the primary cause for forming nonmagnetic Fermi-liquid ground state consisting of heavy quasiparticles, the Ruderman-Kittel-Kasuya-Yosida (RKKY) interaction tends to stabilize a magnetically ordered ground state.  The competition between the Kondo effect and the RKKY interaction leads to notable many-body effect, providing  an ideal platform to study quantum criticality, which can be tuned  by non-thermal parameters such as pressure, chemical substitution, or magnetic field.   

In the vicinity of the quantum critical point (QCP), where the long-range magnetic order vanishes,  novel quantum phenomena are often observed as a result of strong quantum fluctuations.  Of particularly interest is how the heavy quasiparticles evolve near the QCP.  There are two contrasting scenarios for the fate of the heavy quasiparticles at the QCP \cite{Si13,Si14,Wirth16}.  One is the spin-density-wave (SDW) scenario based on spin fluctuations \cite{Moriya73A,Moriya73B,Hertz76,Mills93}.  In this scenario, the magnetic order vanishes continuously without breaking up the heavy quasiparticles at the QCP.  The change of the Fermi surface across the QCP is only due to the SDW gap formation and thus it involves no significant change in the nature of the $c$-$f$ hybridization.  Therefore,  the ``large Fermi surface" consisting of the heavy quasiparticles persists even in the magnetically ordered phase.   Figure\,1(a) illustrates  a schematic phase diagram of the SDW scenario, where $T_N$ and $T^*$ represent  the N\'eel temperature and  the crossover temperature from small to large Fermi surface.   The other  is the Kondo breakdown scenario, as illustrated in Fig.\,1(b) \cite{Si01,Coleman01}.  In this scenario, $T_N$ and $T^*$ simultaneously disappear at the QCP, and consequently the Fermi surface changes abruptly at the QCP.  In the magnetically ordered phase, $f$-electrons are completely localized due to the destruction of Kondo screening.  Thus the heavy quasiparticles disappear on entering the magnetically ordered state, leading to a ``small Fermi surface" state.  This Kondo breakdown  scenario has been extensively discussed in YbRh$_2$Si$_2$ \cite{Gegenwart07}.  
 
CeRhIn$_5$ is a typical heavy fermion compound with an antiferromagnetic  (AFM) ground state; at ambient pressure, CeRhIn$_5$ undergoes a transition to an AFM state with an incommensurate magnetic wave vector {\boldmath $q$}$_M$=(0.5,0.5,0.297) at $T_N = 3.8$\,K.    Pressurizing CeRhIn$_5$ tunes its magnetic transition toward a QCP and induces unconventional superconductivity that coexists with the AFM order at $P\approx$1.75\,GPa \cite{Knebel08}.    A drastic change of the Fermi surface at the QCP under pressure has been reported by de Haas-van Alphen (dHvA) measurements \cite{Shishido05}.  This  has been discussed in terms of the abrupt change of the Fermi surface  from small to large ones at the QCP \cite{Wirth16}.   In addition, at zero pressure, the dHvA oscillations are essentially the same as those of LaRhIn$_5$, suggesting that the Ce 4$f$ electrons may not contribute to the Fermi surface \cite{Shishido02,Harrison04}.  These results have been interpreted as an indication of Kondo breakdown in CeRhIn$_5$, which is supported by dynamical mean field calculations reporting that the $f$-electrons become localized upon entering the magnetic phase \cite{Haule10}.  
On the other hand,  specific heat measurements report that the large electronic specific coefficient $\gamma$  persists even below $T_N$ \cite{Hegger00,Cornelius00}.  This suggests the presence of heavy quasiparticles in the magnetically ordered state, implying a large Fermi surface, which appears to be consistent with the SDW scenario.  
Slave-boson mean-field theory on the basis of the periodic Anderson model for CeRhIn$_5$ was able to explain the dHvA measurements as well as the large $\gamma$ in the AFM phase supporting the SDW scenario  \cite{Watanabe10}. Furthermore, very recent angle-resolved photoemission spectroscopy (ARPES) measurements down to 8\,K observe signs of the Kondo effect although  the Fermi surface appears to be small \cite{Chen18}.  However, these ARPES measurements have been done well above the AFM ordering temperature. Thus, the real ground state, and in particular the fate of the Kondo effect and the heavy quasiparticles in the AFM phase remain controversial in CeRhIn$_5$.

\begin{figure}[t]
	\begin{center}
		\includegraphics[width=1\linewidth]{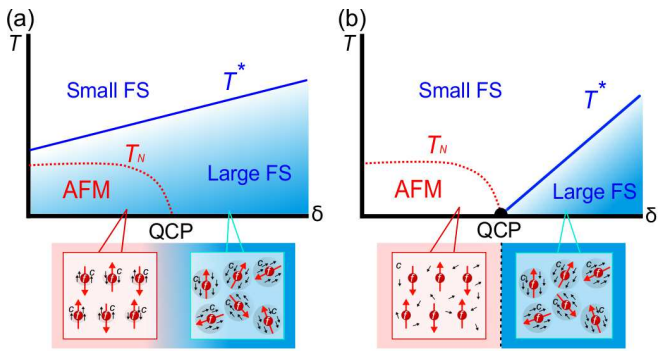}
	\end{center}
	\caption{(a) Schematic phase diagram for the SDW scenario.  $\delta$ is a non-thermal parameter, such as pressure, doping and magnetic field.  $T^*$ and $T_N$ represent the temperature at which heavy quasiparticles are formed and the AFM temperature, respectively. A large Fermi surface is formed even below $T_N$.  The red and black arrows in the lower panels represent the spins of $f$- and conduction electrons, respectively. (b) The same phase diagram for the Kondo breakdown scenario.  A small Fermi surface is formed inside the AFM phase. }
\end{figure}

  The most direct way to resolve this problem is to reveal the low-energy electronic structure near the Fermi level below and above the AFM ordering temperature.    As the local density of states (DOS) is directly measured by scanning tunneling microscopy (STM), it provides indispensable information on the heavy quasiparticles.   In fact, the  energy gap owing to the  hybridization between conduction- and $f$- electrons, $c$-$f$ hybridization, has been reported in  heavy fermion materials, such as CeCoIn$_5$ \cite{Aynajian12,Allan13,Zhou13,Haze18}, URu$_2$Si$_2$ \cite{Aynajian10}, SmB$_6$ \cite{Rossler14}, and YbRh$_2$Si$_2$ \cite{Ernst11}.   
  
 Recently heavy fermion materials, including CeCoIn$_5$ and CeRhIn$_5$, were epitaxially grown by the state-of-the-art molecular beam epitaxy (MBE) technique in ultrahigh vacuum \cite{ShiShido10}.   In this technique, the crystal growth can be controlled at atomic layer level, as revealed by the superconductivity in CeCoIn$_5$/YbCoIn$_5$ and CeCoIn$_5$/CeRhIn$_5$ superlattices which consists of one unit-cell-thick CeCoIn$_5$ \cite{Mizukami11,Goh12,Shimozawa14,Ishii16,Shimozawa16,Naritsuka17}.  Therefore this technique opens up the possibility to produce wide fresh atomically flat terraces, as required for STM measurements\cite{Haze18}.  Here, using  a combined system of MBE and low-temperature ultrahigh vacuum STM, we perform an {\it in situ} STM study of high quality thin films of CeRhIn$_5$.  The STM spectra above $T_N$ clearly resolve the $c$-$f$ hybridization gap and  the crystal electric field excitations.  Below $T_N$,  a small dip structure  in the spectra appears at the Fermi energy.  The spectra below $T_N$  provide the most direct evidence for the presence of itinerant heavy $f$-electrons participating in the Fermi surface even in the magnetically ordered state of CeRhIn$_5$.

\section{experimental}

The $c$-axis oriented epitaxial CeRhIn$_5$ films with tetragonal crystal structure  are grown on the (001) surface of MgF$_2$ (the lattice constant is $\sim0.4625$\,nm.) substrate by MBE  in ultrahigh vacuum ($\sim 10^{-8}$\,Pa).  Ce and In are evaporated from individually controlled Knudsen cells, while Rh  is evaporated from a crucible heated by electron bombardment.  The thickness of the film is 120\,nm and the typical deposition rate is 0.01-0.02\,nm/s.  The $T$-dependence of the resistivity of the thin film is essentially the same as that of bulk single crystals. The residual resistivity of the film is $\rho_0\approx 1\, \mu\Omega$cm, which is larger than that of high-quality single crystals ($\rho_0\approx 0.2\, \mu\Omega$cm).  The nuclear quadrupole resonance spectra and NMR relaxation rate of thin films are essentially unchanged from bulk single crystals \cite{Yamanaka15, Ishida}. The AFM transition temperature  determined by the kink in the resistivity is $T_N=3.5$\,K, which is slightly reduced from the bulk value of 3.8\,K (See Supplemental Information \@Roman{1} \cite{supple}).  This reduction is likely to be due to the strain effect from the substrate because the lattice constant of MgF$_2$ is slightly smaller than that of CeRhIn$_5$.  However we do not find superconducting feature down to 700 mK.  These results demonstrate that the physical properties of thin films are essentially the same as those of bulk single crystals. The STM experiments have been performed with a combined system of MBE and low-temperature  STM. The films are transfered to the STM head, while being kept in ultrahigh vacuum ($\sim 10^{-8}$\,Pa).  All data of the STM measurements are obtained with a PtIr tip.  All conductance spectra are measured by using a lock-in technique with a modulation voltage $V_M$ and a modulation frequency 997\,Hz.

\begin{figure}[t]
	\begin{center}
		\includegraphics[width=1\linewidth]{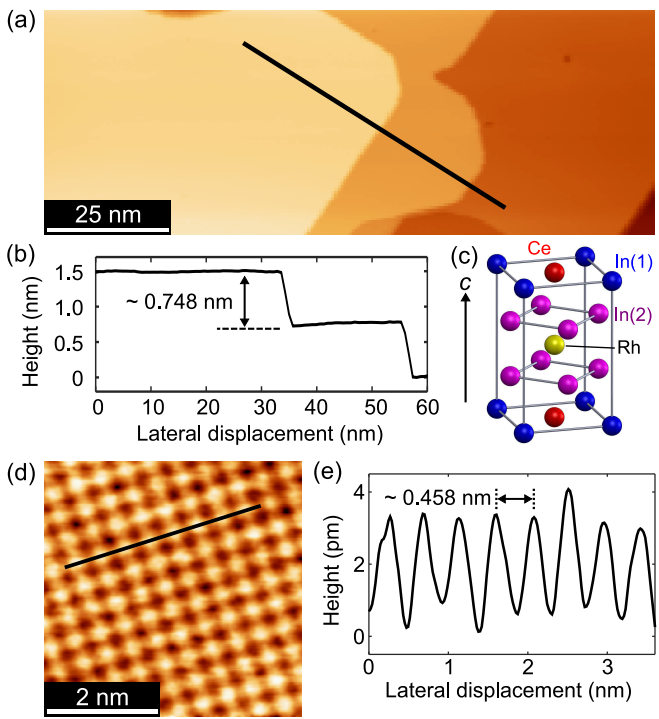}
	\end{center}
	\caption{(a) Topographic STM image of an MBE-grown CeRhIn$_5$ thin film in a wide area.   Tunnel parameters: sample bias voltage $V_s = -500$\,mV and tunneling current $I_T = 70$\,pA.  (b) The cross sectional profile along the black line in (a).  (c) A crystal structure of CeRhIn$_5$.  (d) An atomically resolved topography taken on a flat terrace.  Tunnel parameters: $V_S = -80$\,mV and $I_T = 1$\,nA.  (e) The line profile along the black line in (d).}
\end{figure}

\section{results}

Figures\,2(a) and (b) show a typical topographic image and a corresponding line profile on a CeRhIn$_5$ film, respectively.  Atomically flat terraces, which are much wider than those of cleaved surfaces of a single crystal \cite{Aynajian12},  are clearly resolved.    As shown in Fig.\,2(b), terraces are separated by a step of $\sim$0.748\,nm, which coincides with the lattice constant along the $c$ axis (0.754\,nm \cite{Moshopoulou01}).  This indicates that the whole surface of the thin film consists of the same atomic layer of CeRhIn$_5$ crystal (Fig.\,2(c)).  Figure\,2(d) depicts the STM topograph with atomic resolution of CeRhIn$_5$ thin film.  The cross section along the black line is shown in Fig.\,2(e).   The distance between these spots determined by this image is 0.458\,nm which well coincides with the in-plane lattice constant determined by X-ray diffraction \cite{Moshopoulou01}.  Therefore, the bright spots correspond to Ce  or In atoms in the CeIn(1) layer or to Rh atoms  in the Rh layer. A previous study reports that the top surface is always the CeIn(1) layer in CeCoIn$_5$ thin films grown by MBE \cite{Haze18}.  Moreover, Ce atoms are not clearly resolved in the CeIn(1) layer, as the $p$-orbitals of the In atoms  are well extended perpendicular to the surface, while all orbitals of the Ce atoms are less extended.  Therefore the bright spots in Fig.\,2(c) are most likely to be attributed to In(1) atoms.

\begin{figure}[t]
	\begin{center}
		\includegraphics[width=1\linewidth]{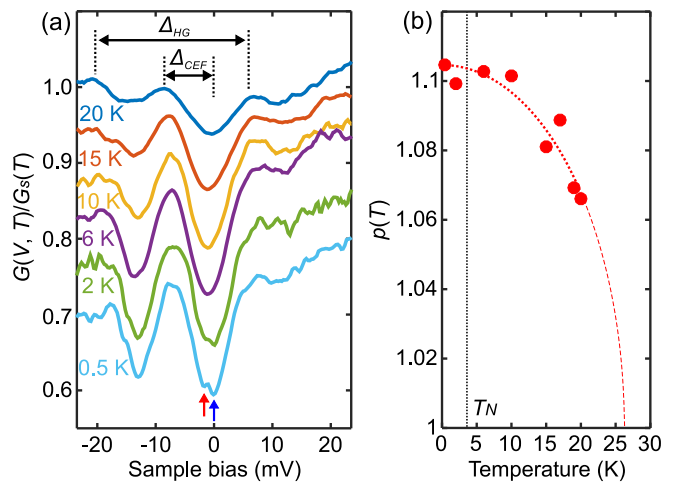}
	\end{center}
	\caption{(a) Tunneling conductance spectra taken at several temperatures.  Tunnel parameters: $V_s = -25$\,mV, $I_T = 1$\,nA, and modulation voltage $V_M = 0.3$\,mV.  The spectra are normalized by each setpoint conductance $G_s(T) \equiv G(V_s,T)$.  The spectra are vertically shifted  for clarity.    (b) Temperature dependence of the ratio of the setpoint conductance to that at $E = -2$\,meV.  The dashed line is a fit to the data. For details, see the text. }
\end{figure}

Figure\,3(a) displays conductance spectra $G(V,T)=dI/dV$ that are proportional to the density of states (DOS)  at several temperatures below 20\,K. We note that this DOS corresponds to the conduction electron bands, as the transfer matrix between the tip and sample is dominated by the orbitals of the In atoms \cite{Aynajian12,Haze18}.    For the comparison of the spectra at different temperature,  $G(V,T)$ are normalized by a setpoint conductance $G_s(T)$ at each temperature.  The conductance spectra are essentially the same in the whole terrace, independent of the atomic sites.   At all temperatures, the DOS  exhibits a characteristic dependence on energy $E$ (or sample bias $V$).  The DOS has two minima at around $E=0$, which corresponds to the Fermi energy, and $E\approx -12$\,meV,  and a peak at $E\approx -8$\,meV.  These features are reproduced by different tips (See Supplemental Information \@Roman{2} \cite{supple}).

We point out that the peak at $E\approx -8$\,meV in the DOS is attributed to the crystal electric field (CEF) excitations with energy $\Delta_{CEF}$.  Inelastic neutron scattering measurements  report CEF excitations of $\sim 7$\,meV from the ground state $\Gamma_7^1$ to the first excited state $\Gamma_7^2$ \cite{Christianson02}.    Although the energy scale of the CEF is comparable in our results, we note that the sign is opposite compared to the neutron scattering and ARPES experiments \cite{Christianson02, Chen18}. This implies that the CEF is modified at the surface. The spectra demonstrate the presence of a large hybridization gap $\Delta_{HG}\approx 25$\,meV at all temperatures.  The double minima structure appears as a result of CEF excitations inside the hybridization gap.  CEF excitations have been reported by STM spectra in YbRh$_2$Si$_2$ at the negative bias voltage, but those are observed outside the hybridization gap \cite{Ernst11}.  Besides the hybridization gap, the spectra taken at $T=2$\,K and $T=0.5$\,K exhibit an additional small dip structure at $E=0$ as shown by the blue arrow. We will discuss this later.

Next we discuss the evolution of the $c$-$f$ hybridization quantitatively.  We note that this temperature evolution cannot be explained by the effect of thermal broadening as discussed in the Supplemental Information \@Roman{3} \cite{supple}.  The magnitude of the $c$-$f$ hybridization is represented by the reduction of the DOS due to the formation of the hybridization gap.  However, there is a large ambiguity in determining the DOS reduction, because of the CEF excitation peak and unknown energy dependence within the hybridization gap.  
Here, we estimate the magnitude of the $c$-$f$ hybridization gap by $G(T)$ at $E=-2$\,meV (red arrow in Fig. 3(a)) normalized by $G_s(T)$, which lies inside the hybridization gap but outside the small dip structure developing at low temperature around $E$ = 0 (blue arrow in Fig. 3(a)).
Figure 3(b) depicts the temperature dependence of $p(T)\equiv 1/[G(-2\,{\rm meV},T)/G_s(T)]$.  
As the temperature is lowered, $p(T)$ increases rapidly  and saturates below $\sim10$\,K.  
If we fit $p(T)$ by simply assuming $p(T)=1+p(0)\sqrt{1-(T/T_0)^2}$ as shown by the red dotted line in Fig.\,3(b), we obtain $T_0\sim 26$\,K.  
Moreover, to verify $T_0$ by a different way, we estimate $T_0$ by fitting a Gaussian function to the measured DOS (See Supplemental Information \@Roman{4} \cite{supple}). In this analysis, we obtain $T_0 \sim 23$\,K. 
Although these estimates of $T_0$,  at which the hybridization gap disappears, should be scrutinized, the obtained $T_0$ in both analyses is close to the Kondo temperature $T_K \sim 25$\,K reported from the neutron scattering, below which the $c$-$f$ hybridization gap formation occurs \cite{Christianson02}.  The observation of the hybridization gap clearly demonstrates the presence of itinerant $f$ electrons forming heavy quasiparticles and participating in the Fermi surface.

Next we discuss the small dip structure around the Fermi energy below $T$ = 2\,K, indicated by the blue arrow in Fig.\,3(a).  We note that this structure is reproduced by different tips (See Supplemental Information \@Roman{2} \cite{supple}).   We point out that this dip appears as a result of the AFM ordering.  In order to confirm the existence of the dip structure, Fig.\,4 shows the conductance spectra below 4\,K in a narrow bias voltage range with high signal to noise ratio.  The spectra are normalized by that at 5\,K $G(V,T)/G(V,5{\,\rm K})$, where the hybridization gap is almost saturated.  Obviously the reduction of the DOS, which is absent at 4\,K, appears at $T$ = 3\,K.  Further reduction of the temperature leads to a further reduction of the DOS.  The inset of Fig.\,4 shows the temperature dependence of the area, which  corresponds to the reduced DOS shown by the filled area in the main panel of Fig.\,4.   The temperature,  at which the reduction of the DOS occurs, is close to  $T_{N}$. 
These results clearly demonstrate the coexistence of the $c$-$f$ hybridization gap and AFM order, providing direct evidence that  the heavy quasiparticles persist even in the AFM ordered phase.

\begin{figure}[t]
	\begin{center}
		\includegraphics[width=0.75\linewidth]{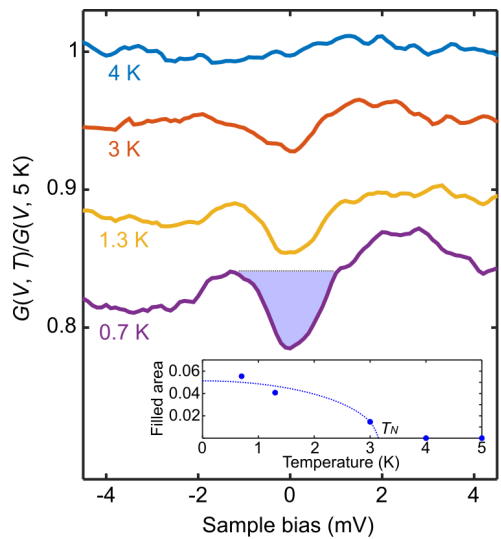}
	\end{center}
	\caption{Tunneling conductance spectra below 4\,K in a narrow energy range around Fermi energy.  Tunnel parameters: $V_s = -5$\,mV, $I_T = 1$\,nA, and $V_M = 0.1$\,mV.    Conductance spectra are normalized by that at 5\,K.  Spectra are vertically shifted for clarity.  The inset shows the temperature dependence of the reduced DOS, as shown by the filled area in the main panel.  The dashed line is guide for eyes.}
\end{figure}

\section{discussion}

Our results demonstrate that the $c$-$f$ hybridization gap of CeRhIn$_5$ thin films is observable at 20\,K, although it has not been observed for cleaved crystals at 20 K \cite{Aynajian12}. This is because the Kondo temperature might be increased by the strain effect from the substrate. Since the lattice constant of the substrate (MgF$_2$) is 0.4625 nm and that of CeRhIn$_5$ is 0.4656 nm, a positive pressure is induced by the substrate. Generally, the $c$-$f$ hybridization is enhanced by this pressure. 
However, the pressure is not as large as to induce superconductivity because we do not find any superconducting feature down to 700 mK (See Supplemental Information \@Roman{1} \cite{supple})). This suggests that although our experiments might agree with cleaved crystals under pressure, we do not cross the quantum critical point and stay in the antiferromagnetic phase of CeRhIn$_5$.  Therefore, our conclusion about the coexistence of the c-f hybridization gap and AFM order in CeRhIn$_5$ is not influenced by the effect of thin films.

The existence of heavy quasiparticles in the AFM ordered state supports the SDW scenario rather than the Kondo breakdown scenario. We note, however, that the fraction of $f$-electrons participating in the formation of a large Fermi surface is difficult to determine.  The ARPES measurements performed at 8\,K indicate that the fraction of $f$-electrons is small, which is supported by thermodynamics indicating a large residual entropy shown by the specific heat at this temperature \cite{Hegger00,Cornelius00}.  On the other hand, the saturation of the hybridization gap seen in our STM measurements, which are done at lower temperature, can be interpreted as  a sign of complete Kondo screening.  This discrepancy may be reconciled by considering that the $f$-electrons split up in two parts; one participating in and one avoiding the Fermi surface.  Thus, although our STM and recent ARPES measurements reveal more and more  the properties of CeRhIn$_5$ at low temperatures, there are still open questions, concerning the fate of the strongly correlated $f$-electrons. That is all the more important as CeRhIn$_5$ is a prototypical heavy fermion compound and was pointed out as a good candidate for the Kondo breakdown scenario.
Finally, we note that in YbRh$_2$Si$_2$, for which the Kondo breakdown scenario has been widely discussed, STM spectra have not been reported below $T_N$.  Therefore, spectra at even lower temperature are required  for comparison with CeRhIn$_5$ and the spectroscopic evidence for the Kondo breakdown scenario.


\section{conclusion}

To clarify the fate of the heavy quasiparticles  in the magnetically ordered phase,  we performed $in$ $situ$ STM measurements down to 500\,mK on a high quality CeRhIn$_5$ thin films grown by MBE.   The hybridization gap as well as the CEF excitations are clearly resolved below 20\,K.  Remarkably,  these spectra are preserved even below the N\'eel temperature, which can be identified by the formation of a small dip structure at the Fermi energy.   These results provide direct evidence that heavy $f$-electrons are itinerant and participate in the Fermi surface even in the magnetically ordered phase.

\begin{acknowledgments}
We thank P. Coleman, X. Dai, T. Hanaguri, S. Kirchner, Y. Kuramoto, K. Miyake, H. Shishido, Q.Si, and S. Watanabe  for fruitful discussions. This work was supported by Grants-in-Aid for Scientific Research (KAKENHI) (Nos.\,JP25220710, JP15H02106, JP15H03688, and JP15H05457), Grants-in-Aid for Scientific Research on Innovative Areas “Topological Materials Science” (Nos.\,JP15H05852 and JP15K21717) from Japan Society for the Promotion of Science (JSPS).
\end{acknowledgments}

\bibliography{STMCeRhIn5_180801}

\setcounter{section}{0}

\newpage

.

\newpage

\section* {Appendix}

\section {Physical properties of thin films}

\begin{figure}[b]
	\begin{center}
		\includegraphics[width=1\linewidth]{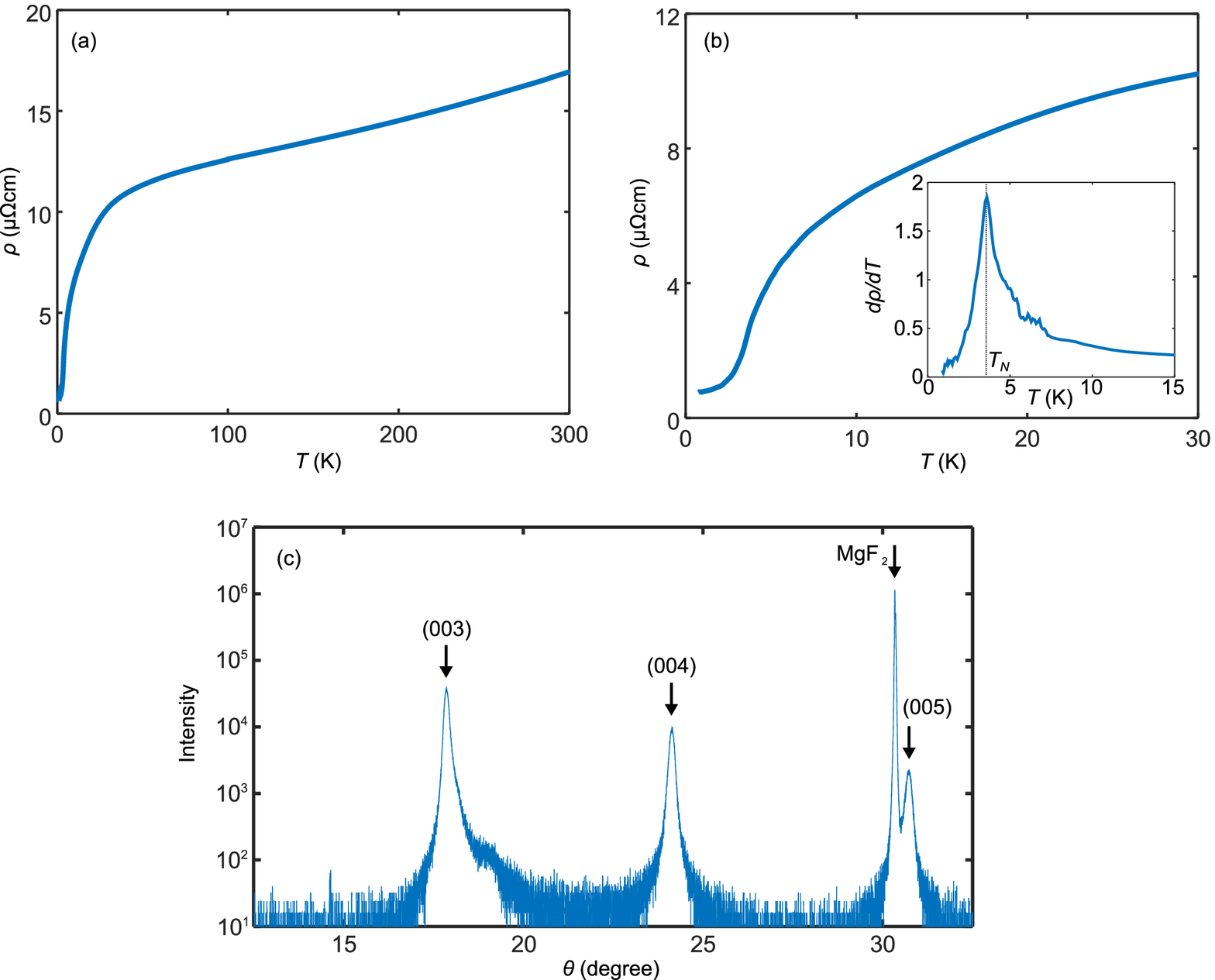}\\
\textbf{FIG.\,S1} \rm{(a) Resistivity measurement of CeRhIn$_5$ thin film in wide temperature range.  (b) Resistivity measurement at low temperature. (Inset) Differential resistivity of the main panel.  (c) X-ray diffraction pattern for CeRhIn$_5$ thin films with a
scattering vector normal to the film plane. Cu-K$\alpha$ radiation is used in the measurement.}
	\end{center}
\end{figure}

In order to demonstrate that the physical properties of a thin film are essentially the same as that of single crystals, Fig.\,S1(a) shows the electrical transport measurement data of a CeRhIn$_5$ thin film taken at the temperature range from 300\,K to 0.7\,K.  To show the transition at low temperature, the data below 30\,K is shown in Fig.\,S1(b). There is a kink structure around 3.5\,K which can be attributed to the AFM transition. This transition is clearly observed as a peak structure in the differential resistivity (see inset in Fig.\,S1(b)).  This transition temperature is slightly reduced from that of single crystals (3.8\,K), which is likely due to the strain effect from the substrate.  However, we do not find any superconducting feature down to 700\,mK, suggesting that the physical properties are not significantly affected by the strain.

To compare the lattice constants of single crystals and thin films, we show the data of X-ray diffraction of thin films in Fig.\,S1(c). The highest peak at around 30 degree corresponds to the lattice of the MgF$_2$ substrate. In addition to the peak of substrate, the (003), (004), and (005) peaks are clearly resolved. We obtain the $c$-axis lattice constant of 0.7541\,nm, which is comparable with that of single crystals (0.7542\,nm).  Therefore, the physical properties of the thin film are essentially the same as those of single crystals.
\\

\section {Reproducibility of the spectra}

STM spectra sometimes show strange structures due to adsorptions to the tip apex. Thus, it is important to show the reproducibility of the spectra using different tips. The characteristic feature in our spectra is the double minima structure at $E=-12$\,meV and $E=0$ as shown in Fig.\,3(a).  In order to show the reproducibility of this feature, Fig.\,S2(a) displays the spectrum taken at 5\,K using a different tip.  Obviously, this spectrum reproduces the double minima structure. Moreover, our spectra show small dip structure around E = 0, whose gap size is around a few meV below the N\'eel transition temperature as shown in Fig.\,4. Figure S2(b) displays the spectrum taken at 0.5\,K. We note that the spectrum is not normalized by any spectrum. In this spectrum, a small dip structure is clearly reproduced although there is a background caused by the hybridization gap. Therefore, we conclude that the obtained spectra are not an artifact but intrinsic features.
\\

\begin{figure}[h]
	\begin{center}
		\includegraphics[width=1\linewidth]{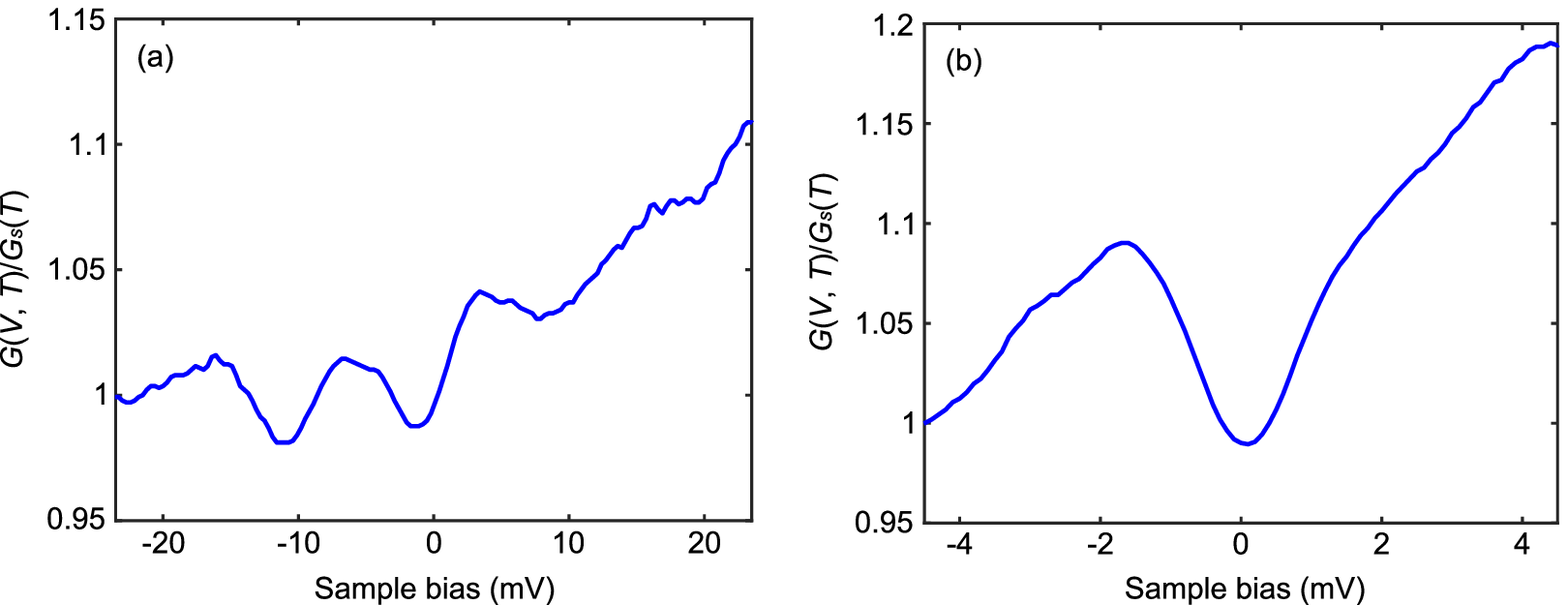}\\
\textbf{FIG.\,S2} \rm{(a) A conductance spectrum taken at 5\,K.  Tunnel parameters: $V_s = -30$\,mV, $I_T = 2$\,nA, and $V_M = 0.3$\,mV.   (b)  A conductance spectrum taken at 0.5\,K.  Tunnel parameters: $V_s = -5$\,mV, $I_T = 1$\,nA, and $V_M = 0.1$\,mV. } 
	\end{center}
\end{figure}

\section {effect of thermal broadening}

In Figure 3, the dips at $E=0$ and $E=-12$\,meV, which are created by the $c$-$f$ hybridization, vanish with increasing temperature and become very shallow at 20\,K.  Here, we show that the conductance spectra at high temperatures in Fig.\,3 cannot be explained only by the thermal broadening.
The blue line in Fig.\,S3 shows the spectrum at 20\,K obtained by using the spectrum at 0.5\,K and taking thermal broadening into account.
The actually observed spectrum at 20\,K is shown by the red line in Fig.S3.  
Obviously, the dip structure in the experimentally measured spectrum is much shallower than the thermal broadened spectrum.\\

\begin{figure}[h]
	\begin{center}
		\includegraphics[width=0.7\linewidth]{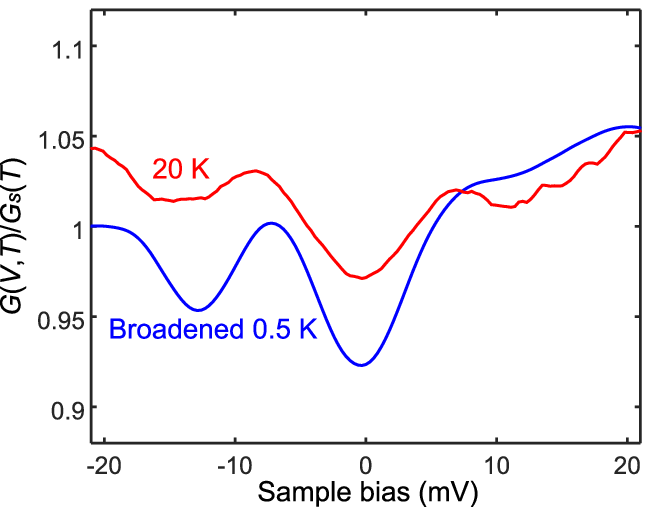}\\
\textbf{FIG.\,S3} \rm{Tunneling conductance spectra taken at 20\,K and that at 0.5\,K broadened by Fermi-Dirac function at 20\,K.  The spectra are vertically shifted for clarity.} 
	\end{center}
\end{figure}

\section {Estimation of the Kondo temperature}

\begin{figure}[h]
	\begin{center}
		\includegraphics[width=1\linewidth]{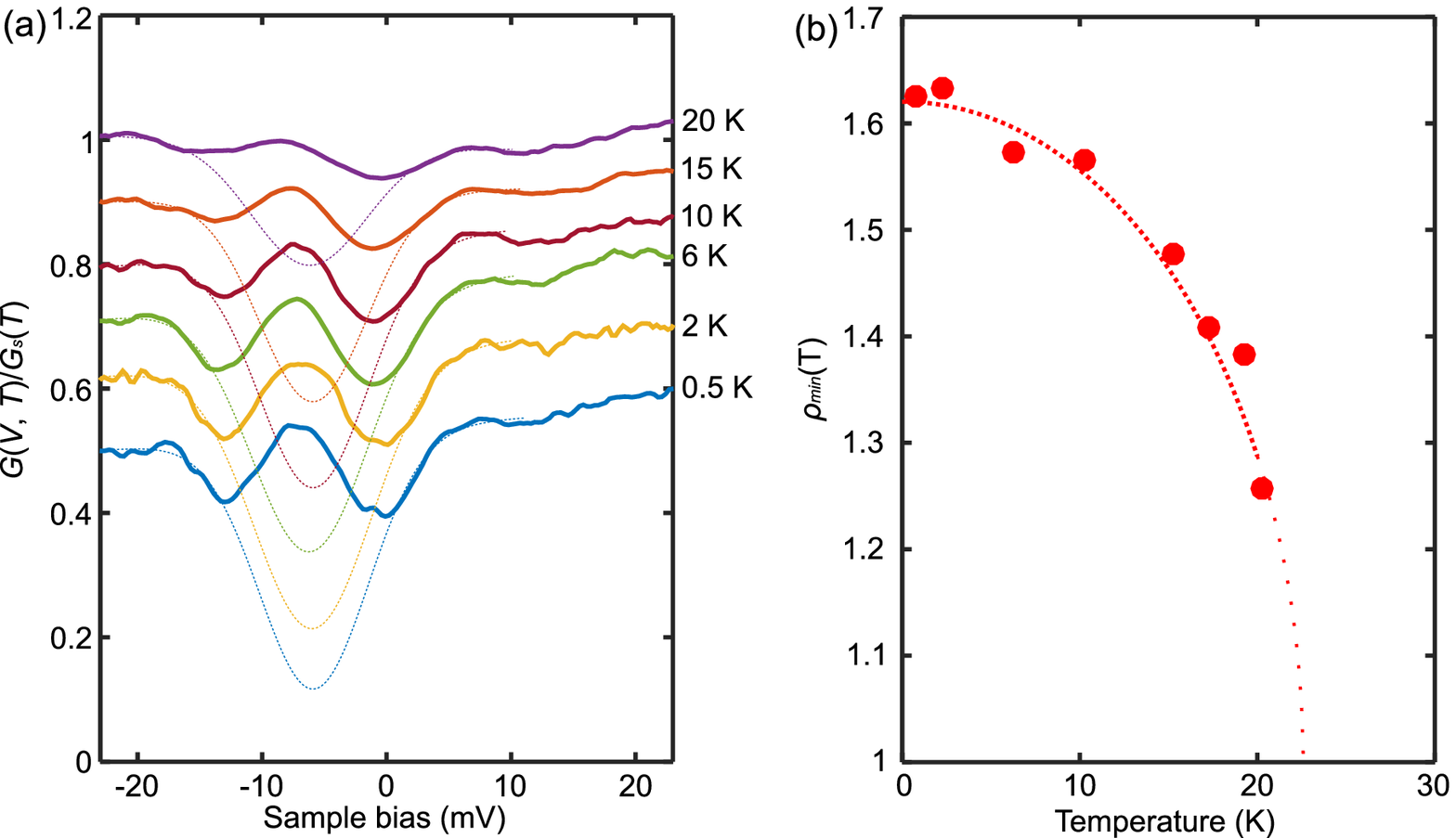}\\
\textbf{FIG.\,S4} \rm{(a) Tunneling conductance spectra taken at several temperatures. The data are the same with those in Fig.\,3.  Dashed lines are fits to the data by a Gaussian function.  (b) Temperature dependence of $\rho_{min} (T)$.  Dotted line is a fit to the data.} 
	\end{center}
\end{figure}

In order to estimate the Kondo temperature by another method than that in Fig.\,3(b), we fit a Gaussian function ignoring the CEF feature to the measured spectrum, as shown in Fig.\,S4(a). The dotted lines show the fitted spectra. For the quantitative analysis of the $c$-$f$ hybridization, we estimate the magnitude of the $c$-$f$ hybridization gap by the minima of the fitted spectra. Figure S4(b) displays the temperature dependence of the inverse of the minima, which is represented by $\rho_{min} (T)$.  The hybridization gap shows a similar temperature dependence to Fig.\,3(b), and saturates at low temperature. If we fit the temperature dependence of $\rho_{min} (T)$ by $\rho_{min} (T)=1+\rho_{min} (0) \sqrt{1-(T/{T_0})^2}$, we obtain $T_0=23$\,K. Although the estimated values vary depending on the analysis methods, both values are comparable with the Kondo temperature $T_K \sim 25$\,K reported from the neutron scattering. Therefore, we conclude that the $c$-$f$ hybridization develops below 23-26\,K. 
\\

\end{document}